**Tuning the optical bandgap in hybrid layered perovskites through variation of alkyl chain length**

Jasmina A. Sichert*[1,2], Annick Hemmerling[3], Carlos Cardenas-Daw[1,2], Alexander S. Urban*[2,4], Jochen Feldmann*[1]

[1] Chair for Photonics and Optoelectronics, Nano-Institute Munich, Department of Physics,
Ludwig-Maximilians-Universität München, Königinstr. 10, 80539 Munich, Germany

[2] Nanosystems Initiative Munich (NIM) and Center for NanoScience (CeNS),
Schellingstr. 4, 80799 Munich, Germany

[3] Department of Chemistry Ludwig-Maximilians-Universität München,
Butenandtstr. 5-13, 81377 Munich, Germany

[4] Nanospectroscopy Group, Nano-Institute Munich, Department of Physics,
Ludwig-Maximilians-Universität München, Königinstr. 10, 80539 Munich, Germany

Abstract:

Recently, layered hybrid perovskites have been attracting huge interest due to a wide range of possible chemical compositions and the resulting tunability of the materials' properties. In this study, we investigate the effect of the chain length of the organic ligands on the optical properties of stacks of two-dimensional perovskite layers consisting of alkylammonium lead iodide $(C_nH_{2n+1}NH_3)_2PbI_4$ with n=4,…,18. Photoluminescence and absorption spectroscopy reveal a blueshift with increasing chain length n including a jump of 110 meV between the n=10 and n=12 ligands due to a change in octahedral tilting. Using X-ray diffraction (XRD) we determine the crystal structure and find the octahedral tilting to be the main cause of this blueshift. However, for very short chain lengths additional effects further reduce the transition energy. Results of effective mass approximation model calculations show good agreement between the expected reduction of transition energy and measured photoluminescence emission wavelength for these samples. This highlights how octahedral tilting plays a major role in determining the optical bandgap and suggests that miniband formation plays only a minor role in this material.

Introduction:

Perovskites possess an intriguing crystal structure showing many interesting properties and effects such as strong optical absorption, ferroelectricity and superconductivity.[1-4] For the three-dimensional (3D) halide variant of the perovskites the general formula of the crystal structure is $ABX_3$, where A and B are monovalent and divalent cations, respectively, and X is a monovalent anion. The $B^{2+}$ and $X^-$ ions form octahedra, with neighboring octahedra sharing corners, consequently spanning a 3D network.[5, 6] The $A^+$ ion is located in the gaps between 8 such corner-sharing $[BX]^4$ octahedra. The crystal structure allows for phase transitions, cation displacement, distortions, and octahedral tilt that alter the properties of the material.[1, 5, 7-9] In addition to the 3D perovskites there are several other classes of perovskites, amongst them organic-inorganic layered two-dimensional (2D) perovskites.[10-15] In this group, the organic $A^+$ ion is too big to fit inside the interoctahedral space, thus prohibiting the formation of the 3D structure. Instead, these perovskites form sheets of corner-sharing octahedra separated by an organic moiety comprised of the organic ligand. The corresponding crystal structure is then given by the chemical formula $A_2BX_4$. Already in the 1990s, layered perovskites were used for thin-film field effect transistors.[16, 17] Recently, layered perovskites have received significant attention, for example due to their ability to shield and thus increase the stability of 3D perovskite films for solar photovoltaic applications.[18, 19]

In this study, the general formula of the investigated 2D perovskites is $[(R-NH_3)_2PbI_4]$, where R is an organic alkyl chain. The correct chemical formula $(C_nH_{2n+1}NH_3)_2PbI_4$ will hereafter be abbreviated as $C_nPbI_4$ where n is the number of C-atoms. These structures have long been known to naturally form electronic potentials as known for multiple quantum wells.[11, 12, 20] Figure 1a illustrates the chemical structures of these organic-based perovskite multi-layers. The corner-sharing $PbI_6$ form 2D monolayers constituting the quantum wells with the organic ligands forming barriers, separating the quantum wells. In the past, several studies have shown experimentally and theoretically that in 2D perovskites the quantum confinement of the electron and hole and thus the emission wavelength can be tuned by varying the thickness of the inorganic part, so effectively the quantum well thickness.[21-23] In contrast, the present study focuses on the systematic variation of the "barrier" length ($L_B$) achieved through variation of the organic ligand length and its effect on the optical properties of these layered organic-based perovskites. Herein we synthesize films of such multiple stacked quantum wells. The PL emission wavelength shows a consistent blueshift with increasing chain length interrupted by a large jump in the blueshift between the $C_{10}$ and $C_{12}$ ligand-based samples. As all samples are found in an orthorhombic crystal configuration, structural phase transitions can be excluded and octahedral tilting can be isolated as a crucial factor determining the emission wavelength. Additionally, remaining discrepancies of optical measurements and tilting angles indicate that for short ligands additional effects further lower the PL energy. Effective mass approximation (EMA) model calculations suggest that for short ligands, n<12, $L_B$ is short enough to allow electronic coupling between layers. The resulting miniband formation leads to the observed decrease of the PL energy.

Results and discussion:

For the sample preparation a 1.0 molar $C_n$-ammonium iodide solution in dimethylformamide (DMF) and 0.25 molar lead iodide solution in DMF were mixed in a stoichiometric ratio of 2:1 and were drop casted on a glass substrate. The perovskite crystals formed upon evaporation of the organic solvent. Details can be found in the materials and methods section. For a first structural analysis X-ray diffraction (XRD) measurements were carried out (Figure 1b). The analysis of the peaks showed that all samples crystallized in an orthorhombic crystal configuration at room temperature with dominant peaks in the $(002l)$ direction where $l = (1,2,3 ... )$.[24-26] As can be seen in the figure the peak to peak distance decreased with increasing organic molecule length. Using Bragg's law, we determined the octahedron-to-octahedron center-to-center distance between neighboring layers, hereafter referred to as $d_{Bragg}$. It was found to increase linearly from 13.Å for $C_4PbI_4$ to 31.5Å for $C_{18}PbI_4$, in excellent agreement with previously reported values (Figure 1c).[24-26] Using the width of the perovskite layer $d = 6.0$Å, the thickness of the organic ligand layer separating the individual perovskite layers lies between 7.6Å ($C_4PbI_4$) and 25.5Å ($C_{18}PbI_4$).

The color of all films was yellow to orange under ambient illumination at room temperature in agreement with previous reports on perovskite monolayers and nanoplatelets.[23-26] To determine whether the ligand had an effect on the optical properties, we recorded absorption and PL spectra for all samples (Figure 2a). All absorption spectra show the characteristic features of a 2D semiconductor, with a prominent 1s exciton peak, and a step-like continuum absorption at shorter wavelengths. The PL spectra of all samples show a single emission peak, with a small Stokes shift of 14 – 33 meV, indicating excitonic recombination. The PL maximum lies around 521 nm for the films with the shorter hydrocarbon chains (n = 4-10), with a slight blue-shift of 2 nm (9 meV) total occurring with increasing ligand length. The PL emission exhibits a blueshift jump of 23 nm (110 meV) between the $C_{10}PbI_4$ and $C_{12}PbI_4$ samples, followed by a further gradual blueshift of 5 nm (25 meV) up to the $C_{18}PbI_4$ perovskite sample. For detailed values see Table S1 in the supporting information. Similar jumps in PL and absorption are known from layered hybrid perovskites, which is however induced by a phase transition at high temperatures.[27] As shown through the XRD analysis (Figure 1b), all samples here crystallize in the orthorhombic structure, and so a crystal phase transition cannot explain the observed blueshift. However, there can also be structural changes that do not require a phase change, e.g. through a change in the bond angles within the crystal. A correlation between the photoluminescence energy maxima $E_{PL}$ and bond angles was experimentally observed for related layered perovskite structures.[28-34] Indeed, a jump in Pb-I-Pb bridging angles, hereafter referred to as octahedral tilting, for $C_nPbI_4$ with n varying between 4 and 18 was observed by Billing and Lemmerer[24-26] in single crystal X-ray diffraction experiments at the same position as the here observed jump in $E_{PL}$. This can be seen by plotting both the octahedral tilt angles, which were obtained from these publications[24-26], and $E_{PL}$ against the organic ligand chain length (Figure 2c). While the jump aligns nicely and a similar trend can be observed for the longer ligands, the shorter ligands deviate considerably.

This can be visualized more effectively by plotting the tilt angle directly versus $E_{PL}$ (Figure 3a). A linear fit is applied to this graph, with a general correlation between the two parameters: the larger the octahedral tilt, the higher the emission energy (Figure 3b). Again, for longer chain lengths, the data points arrange nicely on the line, while the shorter ones deviate further from it, indicating an additional effect taking place when the perovskite layers come closer together. The data points of $C_4PbI_4$ and $C_6PbI_4$ lie below the fit while $C_8PbI_4$ and $C_{10}PbI_4$ are above the fit.

Already in the 1990s Mitzi and coworkers proposed that the inorganic perovskite layers can be seen as semiconductor quantum wells, with potential wells for both electrons and holes, as confirmed by the absorption spectra. However, one must consider that these quantum wells are extremely thin, at only approximately 6 Å. This leads to a strong blueshift of the energetic bandgap and also to higher exciton binding energies.[21, 22, 36] These latter values turn out to be larger than can be expected simply through reduction in dimension. As the exciton is confined in the 2D inorganic layer, the dimension is reduced which leads to an increase of the exciton binding energy of four times the values of the bulk counterpart.[22] In addition to this reduction in dimension, dielectric confinement, caused by a significantly lower dielectric constant of the organic moiety compared to the inorganic layer, results in a further increase of the exciton binding energy as not all of the Coulomb interaction is strongly screened by the perovskite layer.[11] This leads to binding energies of more than 10 times those of the bulk materials.[11, 23] However, if the perovskite quantum wells are brought close enough together, according to the multiple quantum well picture one can expect an electronic coupling between neighboring quantum wells and can no longer consider them as isolated. In classic semiconductor systems, such as GaAs/AlAs, electronic coupling between quantum wells or *superlattice* formation, is a well-known effect.[37, 38, 39] In contrast to an isolated semiconductor quantum well where the energy eigenvalues are discrete and the density of states are step-function like (Figure 4a), in a superlattice electronic coupling between the quantum wells leads to the formation of a miniband of width 2Δ (Figure 4b), which increases with decreasing quantum well separation. Although the density of states is continuous in the range of the miniband, the charge carriers recombine between the miniband edges. This leads to a transition energy which is smaller than that of an isolated quantum well. In the limit of an infinitely long barrier, the energy levels of the isolated quantum well are recovered. The occurrence of coupling depends on the quantum well potential, effective masses of carriers, quantum well thickness and barrier length. SEM images of the films used in this work confirmed the formation of continuous films with lateral grain sizes varying between several hundreds of nanometers up to micrometers (Figure S1).

In the aforementioned model, the separate dispersion for electrons and holes can be written as:

$$\cos(qL^{QW})\cosh(\alpha L^B) + \frac{1}{2}(\eta - \frac{1}{\eta})\sin(qL^{QW})\sinh(\alpha L^B) = \cos(\delta)$$

with:

$$q = \sqrt{\frac{2m_{e(h)}^{QW} E_{e(h)}}{\hbar^2}}$$

$$\alpha = \sqrt{\frac{2m_{e(h)}^{B} (V_{CB(VB)} - E_{e(h)})}{\hbar^2}}$$

$$\eta = \frac{\alpha m_{e(h)}^{QW}}{q m_{e(h)}^{B}}$$

and with the quantization of electron and hole energies $E_{e(h)}$, the width of the QW and barrier $L_{QW}$ and $L_B$, $m_{e(h)}^{QW}$ and $m_{e(h)}^{B}$ and $\delta$ a real parameter, equal to $q(L_{QW} + L_B)$ where $q((2\pi n)/\sim L)$ with $\sim L$ the length of the lattice. Parameters were set as $L_{QW}$ = 6 Å, $V_{CB}$ = 0.9 eV, $V_{VB}$ = 2.9 eV.[22] $L_B$, as mentioned before, is calculated by subtracting $L_{QW}$ from the Bragg distance, $d_{Bragg}$ - $L_{QW}$. The choice of effective masses is difficult due to a lack of experimental values, especially for the 2D case.[43] Yet, one can expect the values to be significantly larger (up to 3 to 4 times) than in the bulk case,[44] where the effective masses for holes are likely quite similar and approximated to be between $0.1m_e$ and $0.2m_e$.[45] Hence, we have chosen $m_{QW}$* = 0.5$m_e$ and $m_B$* = 1$m_e$, which should constitute a good approximation of the situation found in our case.

The modified band gap energy $E_g^{SL}$ is given by the formula:

$$E_g^{SL}(n) = E_g^{3D} + E_e(n, q = 0) + E_h(n, q = 0)$$

where $E_g^{3D}$ is the bandgap energy of the 3D material. The calculated values for $E_g^{SL}$ versus ligand length are depicted in Figure 4c. The energy assumes a constant value $E_g^{SL}$=2.68 eV for all barrier widths $L_B$ from an infinite length down to the equivalent of a ligand length of n=12. For shorter ligand lengths we start to see a decrease in the energy, with an exponential dependence down to a value of 2.65 eV for the lowest ligand length applied (n=4). For infinite barrier spacings, the Kronig-Penney yields the energy $E_g^{QW}$ of an isolated quantum well.

$$E_g^{QW} = E_g^{3D} + E_e(n \to \infty, q = 0) + E_h(n \to \infty, q = 0)$$

The model calculations thus show that for long ligands (n>11) there is no miniband formation. For shorter ligands, miniband formation occurs, lowering the energy by $\Delta E(n) = E_g^{QW} - E_g^{SL}(n)$ between 2 meV and 28 meV for the ligand lengths of n=10 and n=4, respectively. For detailed values, see Table S2 of the Supporting Information. These results agree with the findings by Even et al.,[43] who predicted miniband formation for n<12 using lower effective masses (0.3 $m_e$ and 0.75 $m_e$) and of Ahmad et al. who reported no interlayer coupling for ligand lengths of n=12.[46]

While the trend of our values agrees well with predicted values, the absolute values are still off by a significant margin (2.65 eV - 2.68 eV for our values compared with 2.38 eV - 2.52 eV for $E_{PL}$ as obtained in our experiments). This total discrepancy of between 160 meV and 270 meV can be explained by the fact that we are comparing transition energies from PL measurements, for which one also has to consider the exciton binding energy which can reach values in the hundreds of meVs, far greater than the bulk values.[22, 23, 47] Experimentally, a value of 220 meV for $E_B$ in $C_{10}PbI_4$ was found, in good agreement with our data and proposed model.[20]

These calculations show that for all samples with ligands shorter than n=12 the transition energies do not only depend on the tilting angle but additional effects do play a role. According to effective mass approximation (EMA) model calculations, miniband formation could explain these effects. In order to see the sole dependence of the tilting angle on the PL emission energy, we added the calculated energy drop due to miniband formation ΔE(n) to the experimentally determined energies for the excitonic PL. These artificially decoupled PL energy values are replotted as a function of the octahedral tilt (Figure 4d). Taking this into consideration, nearly all points lie on a straight line nicely showing a linear dependence of the tilting angle on the transition energy and at the same time strongly suggesting that the observed deviations from this linearity as seen in Figure 3a for n<12 are due to miniband formation. One must note that Even et al., suggest that the EMA approach might be invalid for ultrathin perovskite layers due to the non-parabolicity of the VB and CB.[43] Accordingly, it was suggested that density functional theory (DFT) calculations would be necessary to accurately describe these systems. However, in a recent publication by Traore et al.[35] the data obtained from DFT calculations still show a mismatch with the optical data for the shortest ligands used. Thus, the tilting angle does not seem sufficient to explain the observed shift in the PL emission data. Consequently, we conclude that the transition energies of monolayer perovskites are determined by two factors, the octahedral tilt angle and the degree of miniband formation resulting from electronic coupling between the quantum wells. While only a small effect due to the high effective masses of the charge carriers, it is currently the only proposed model that can explain the deviations observed for the systems comprising the shortest alkyl chains.

Summary:

In summary, we have investigated the optical transition energies in thin films of inorganic-organic multiple perovskite layers based on $(C_nH_{2n+1}NH_3)_2PbI_4$ with the length of the $C_nPbI_4$ varying from $C_4$ to $C_{18}$. All samples were found in an orthorhombic crystal structure conformation, with only a variation in the octahedral tilt angle occurring, dependent on the organic chain length. UV-Vis and PL spectroscopy revealed a blueshift of the optical transition energies with increasing ligand length, with a large jump occurring between the $C_{10}PbI_4$ and $C_{12}PbI_4$ samples. Through structure-function analysis we could identify the octahedral tilt as the dominant factor in this observed shift. Interestingly, for shorter ligands (n<12) electronic coupling between the perovskite layers likely occurs. The resulting miniband formation is suggested to be the cause of a further reduction of the transition energy by values of up to 28 meV for the $C_4PbI_4$ sample. These experimental findings highlight the intricate and exciting nature of layered halide perovskite structures and suggest that more studies are necessary to fully understand and apply these highly-complex systems.

Supplementary Online Material:

Additional Material includes the complete data for PL measurements as well as crystal structures and octahedral tilt angles and observed energy shifts from the expected linear trend. Scanning electron microscopy (SEM) images of $C_{10}PbI_4$ and $C_{12}PbI_4$ samples.

Acknowledgement:

We would like to thank Dr. Bert Nickel and Janina Römer for helpful discussions regarding X-ray diffraction (XRD). This research work was supported by the Bavarian State Ministry of Science, Research, and Arts through the grant "Solar Technologies go Hybrid (SolTech)," and by LMU Munich's Institutional Strategy LMU excellent within the framework of the German Excellence Initiative (A.S.U.).

Materials and Methods:

**Materials:**

First, the precursor salts were obtained by reacting the long-chain amine with HI in slight access to ensure full protonation (ratio of 1.1:1.0). Subsequently, the ammonium salts were dried in the rotary evaporator and washed three times with ethanol. The product were light yellow to white precursor salts. These precursors were dissolved in DMF such that a 1 molar solution was obtained. PbI$_2$ was dissolved in DMF that a 0.25 molar solution was obtained.

Small amounts of the precursor stock solution were mixed in a stoichiometric ratio PbI$_2$ to n-ammonium iodide 1:2. This mixture was heated to 70°C in an oil bath and drop casted on a glass substrate. The samples were placed on a heat plate at 70°C to ensure full evaporation of the solvent.

**Structural analysis:**

XRD analysis were performed on a *Bruker D8 Advance A25* diffractometer using Cu-Kα (λ=1.52 Å) radiation. The powder diffraction pattern was scanned over the angular range of 2 to 60 (2θ) with a step size of 0.05, at room temperature. The database *Mercury*[10] with entries from references[24-26] was used for data analysis.

For SEM measurements, a *Gemini Ultra Plus* field emission scanning electron microscope with a nominal resolution of ~2 nm (*Zeiss*, Germany) was used. The images were collected by the in-lens detector at an electron accelerating voltage of 0.5 kV and a working distance of 1 mm.

**Optical characterization:**

UV-vis absorption was measured with a *Varian Cary 5000* UV-vis-NIR spectrometer. For photoluminescence (PL) measurements the samples were excited with a monochromated Xe-lamp. PL spectra were taken with a *Fluorolog-3 FL3-22* (*Horiba Jobin Yvon GmbH*) spectrometer equipped with a water-cooled *R928 PMT* photomultiplier tube mounted at a 90° angle. The excitation wavelength was set to 400nm, excitation and emission slits to 2 nm, and the step size to 0.5 nm. Comparison with other measurements with slit size 1 nm let us conclude that the error is less than +/- 1nm, *i.e.* less than 6 meV.

**Figure 1.** (a) Sketches of $C_nPbI_4$ crystal structures as reported in references with the interlayer spacing given by $d_{Bragg}$.[24-26] (b) XRD spectra of the monolayer samples. All samples show an orthorhombic crystal configuration. [24-26] (c) Dependence of the interlayer spacing $d_{Bragg}$ derived using Bragg's law from experimental data on the length of the organic ligand chain length. The black line is a linear fit, yielding $d_{Bragg}$ = (1.31*n* + 8.27) Å with *n* the number of C-atoms in the hydrocarbon chains.

**Figure 2.** (a) PL and absorption spectra of the monolayer films at room temperature. (b) Schematic depicting the $PbI_6$ octahedral tilt angle θ. (c) $E_{PL}$ extracted from data in panel a) and $PbI_6$ octahedral tilt angle extracted from literature [24-26] in dependence on the hydrocarbon chain length. There is a large jump for both $E_{PL}$ and θ between n10 and n12.

**Figure 3.** (a) $E_{PL}$ extracted from data in Figure 2a versus $PbI_6$ octahedral tilt angle extracted from literature. [24-26]. The blue line represents a linear fit with a $R^2$-value of 0.9744. While the fit suggests a very strong correlation of $E_{PL}$ and octahedral tilt within the group of long ligands, there are significant deviations for the very short ligands. (b) PL spectra and schematic illustration of the octahedral tilt for $C_4PbI_4$ and $C_{18}PbI_4$.

**Figure 4.** (a) Schematic representation of the energy levels in an isolated quantum well and the corresponding step-like density of electronic states. (b) Schematic representation of the energy levels in a quantum well-superlattice and the corresponding density of electronic states. Electronic coupling leads to the formation of minibands with a width of 2Δ and a concomitant lowering of the superlattice band gap energy $E_g^{SL}$. (c) Kronig-Penney model calculations predicting a decrease of the superlattice band gap energy $E_g^{SL}$ due to miniband formation for short ligands with n=4 - 11. The decrease $\Delta E(n) = E_g^{QW} - E_g^{SL}(n)$ reaches a maximum value of 28 meV for the $C_4PbI_4$ sample. For all samples with n>10 there is no miniband formation and $E_g^{SL}$ remains constant. (d) In order to see the sole dependence of the tilting angle on the PL emission energy, we added the calculated energy drop due to miniband formation ΔE(n) to the experimentally determined energies for the excitonic PL. The result exhibits a much stronger correlation with the octahedral tilt angle than the pure $E_{PL}$ values with a $R^2$-value of 0.9952. This demonstrates that the two effects, octahedral tilting and miniband formation, combined allow a very accurate explanation of the observed $E_{PL}$ of the $C_nPbI_4$ monolayer stacks.

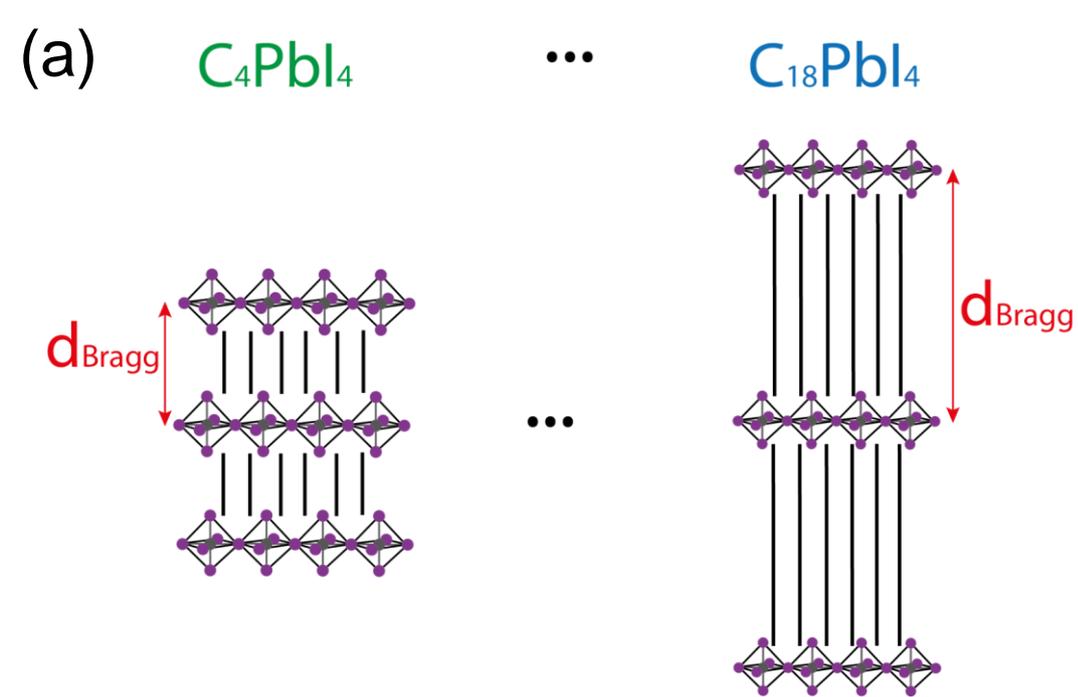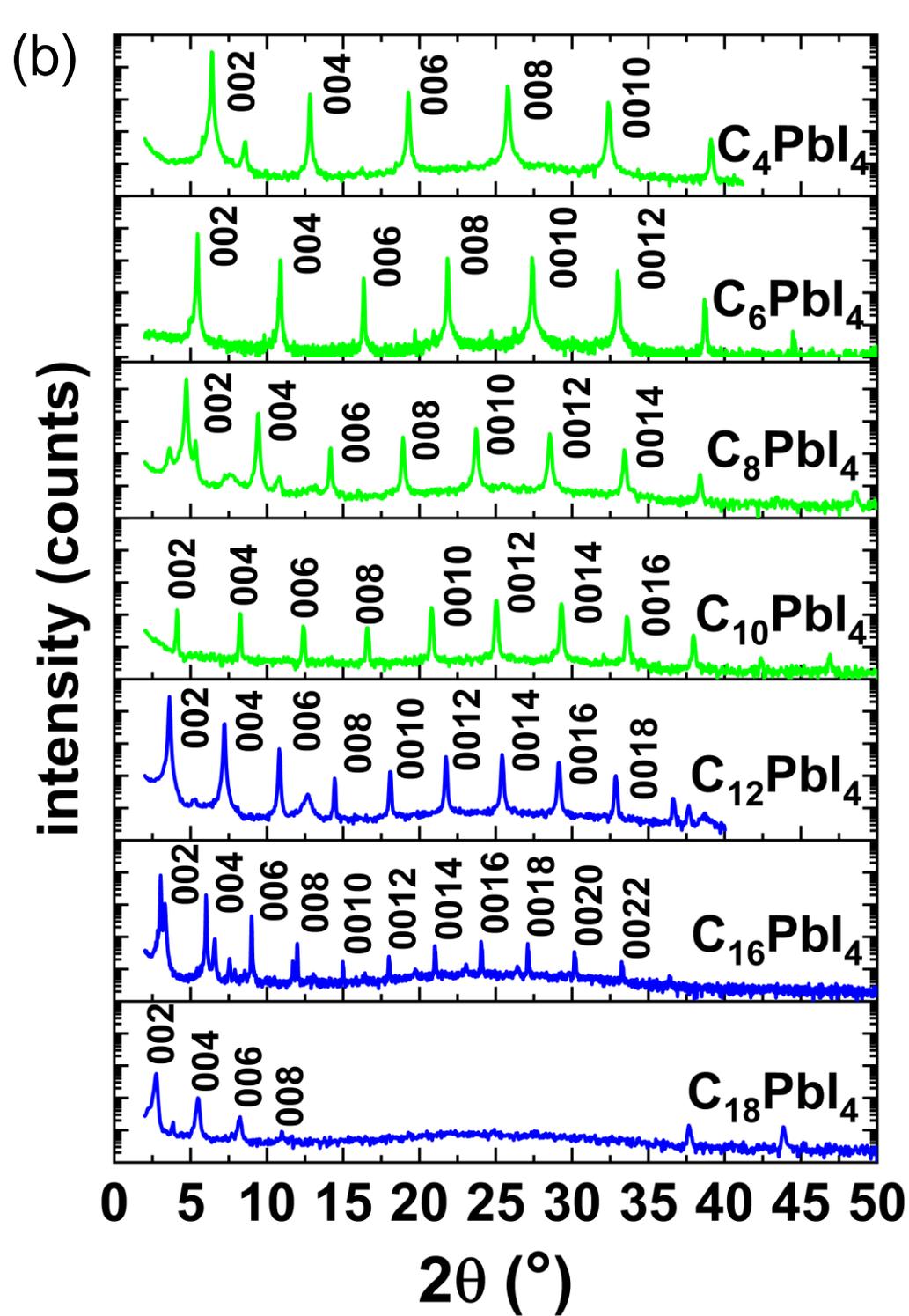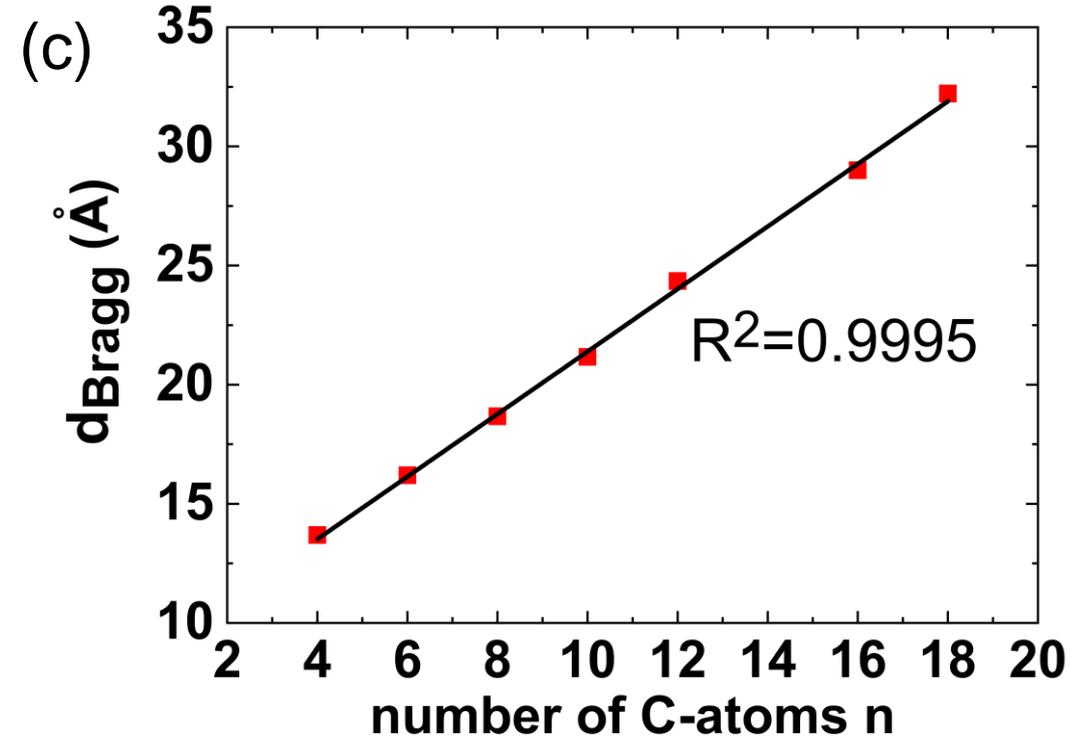

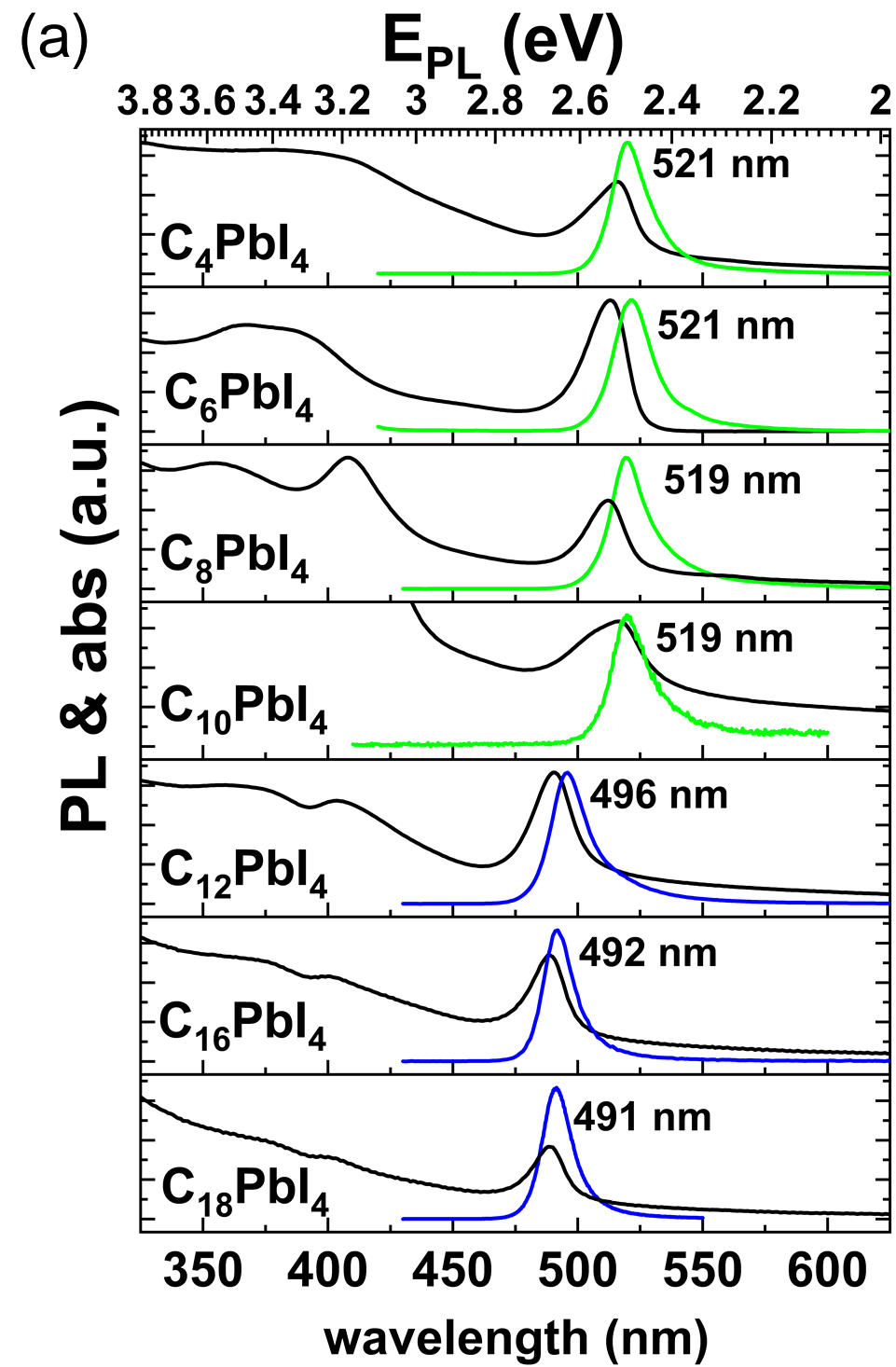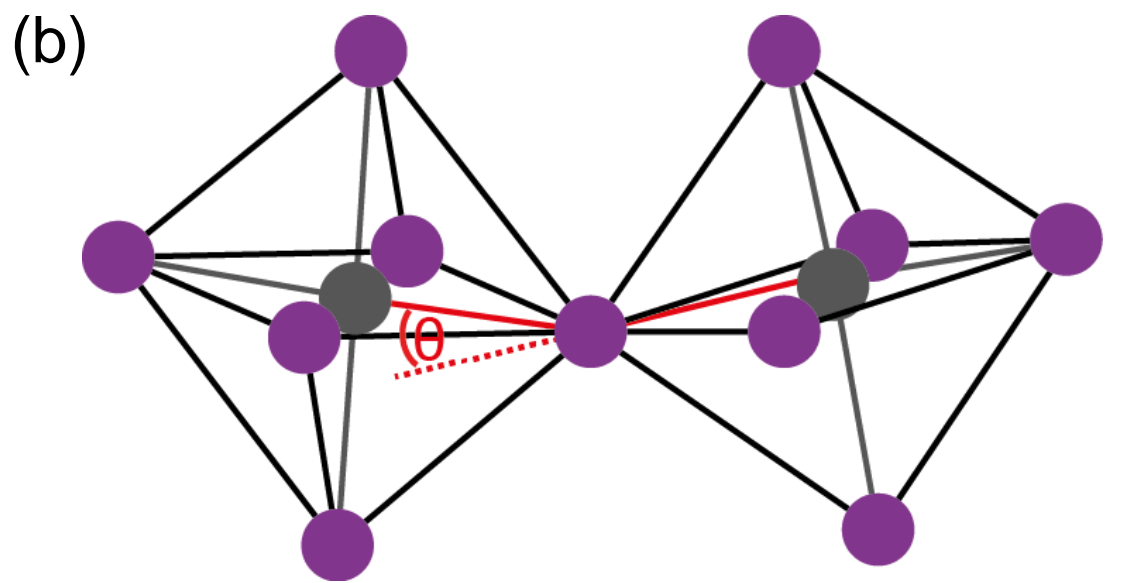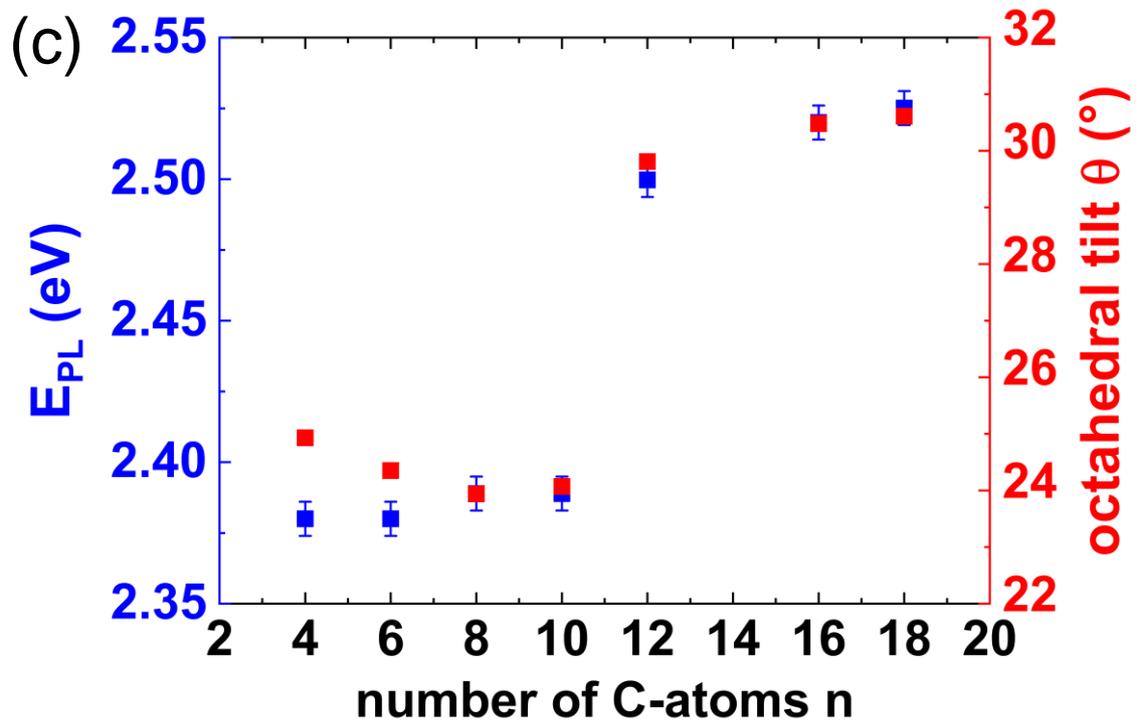

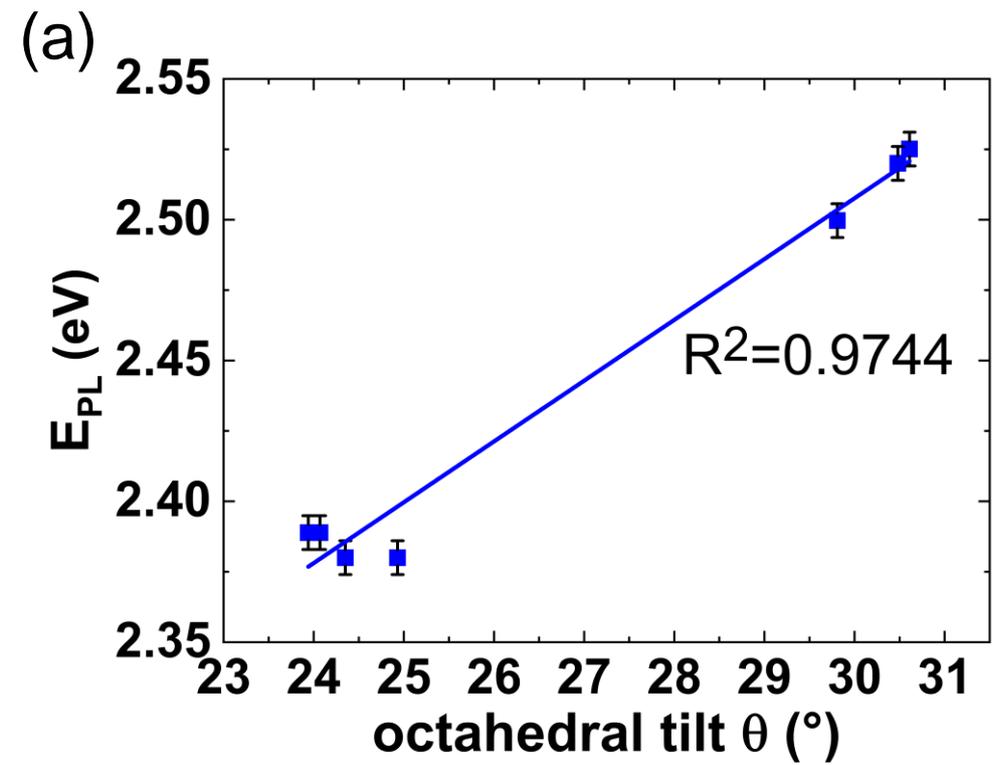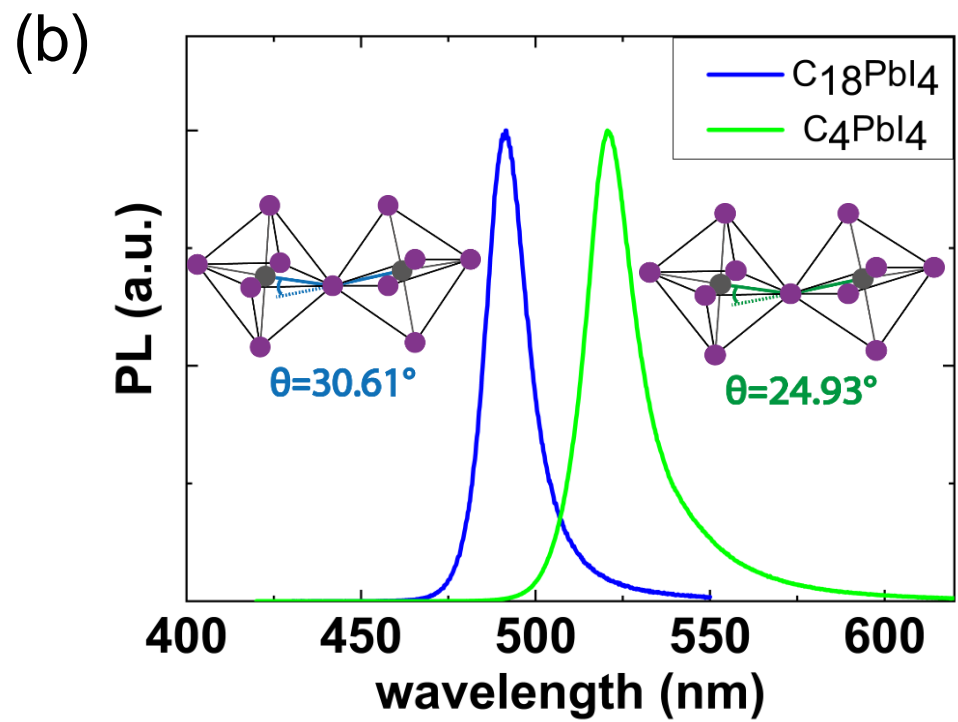

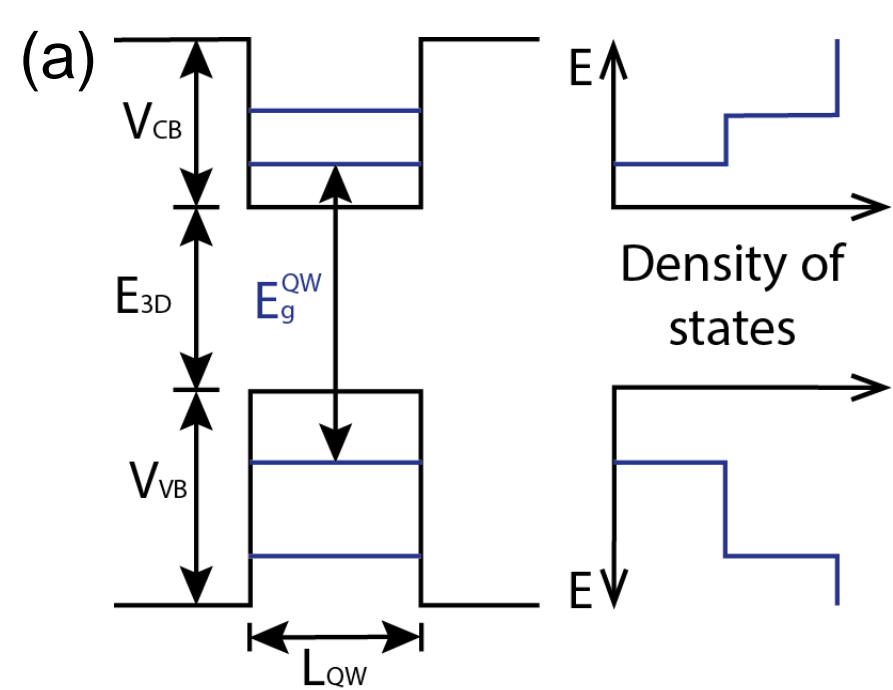
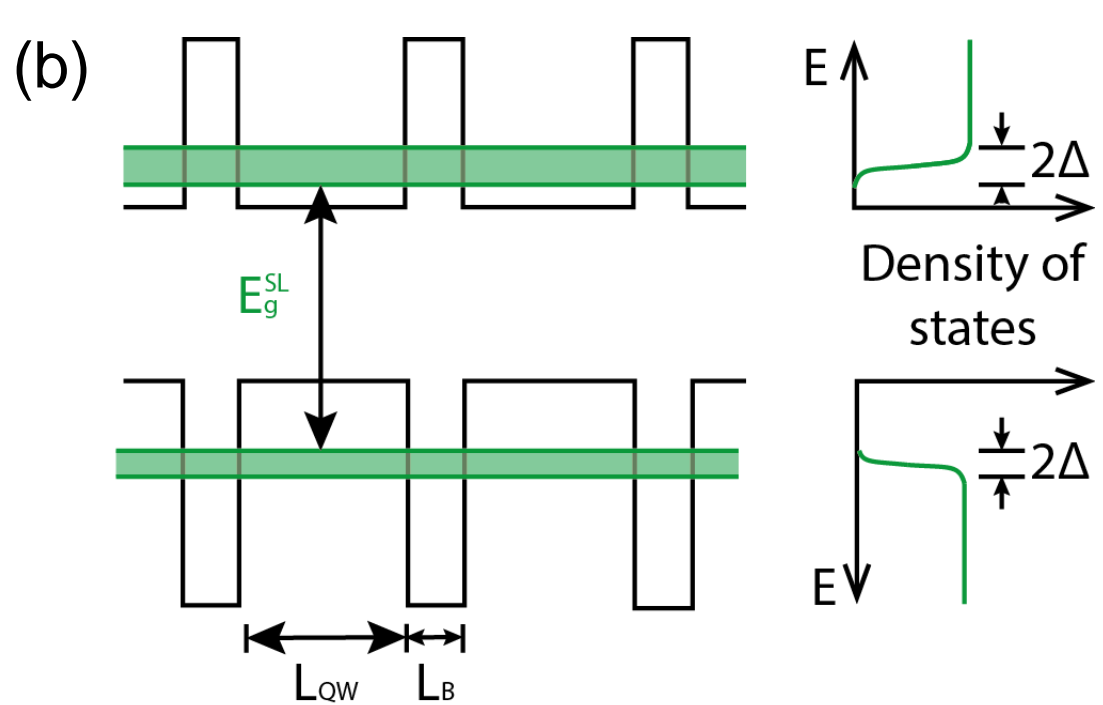
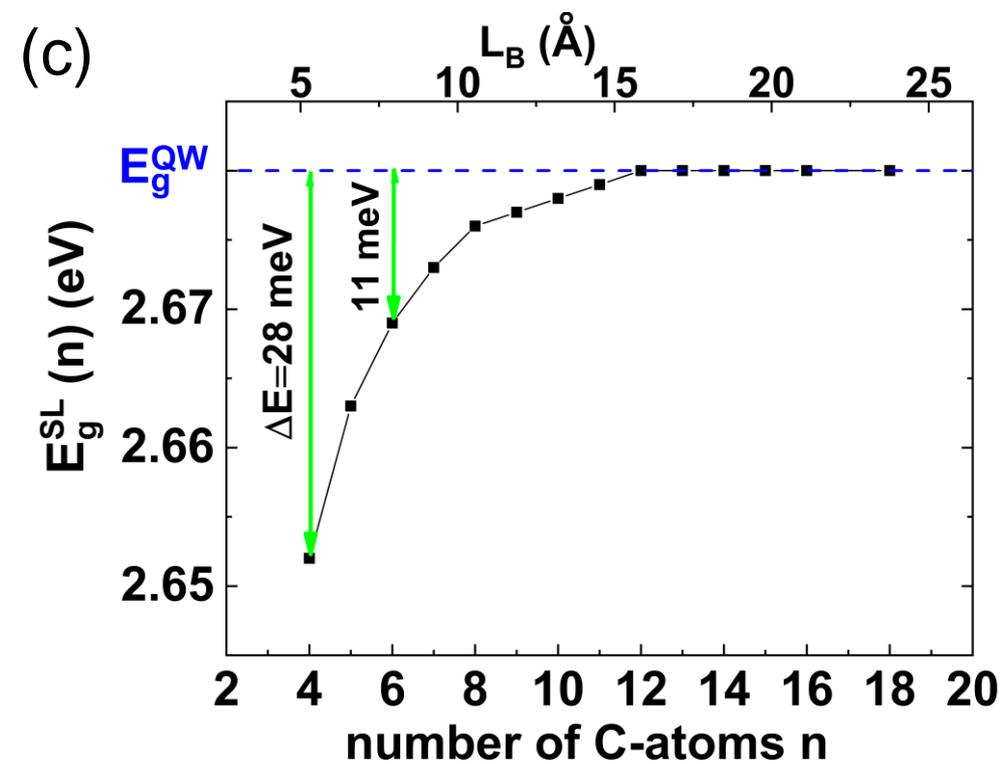
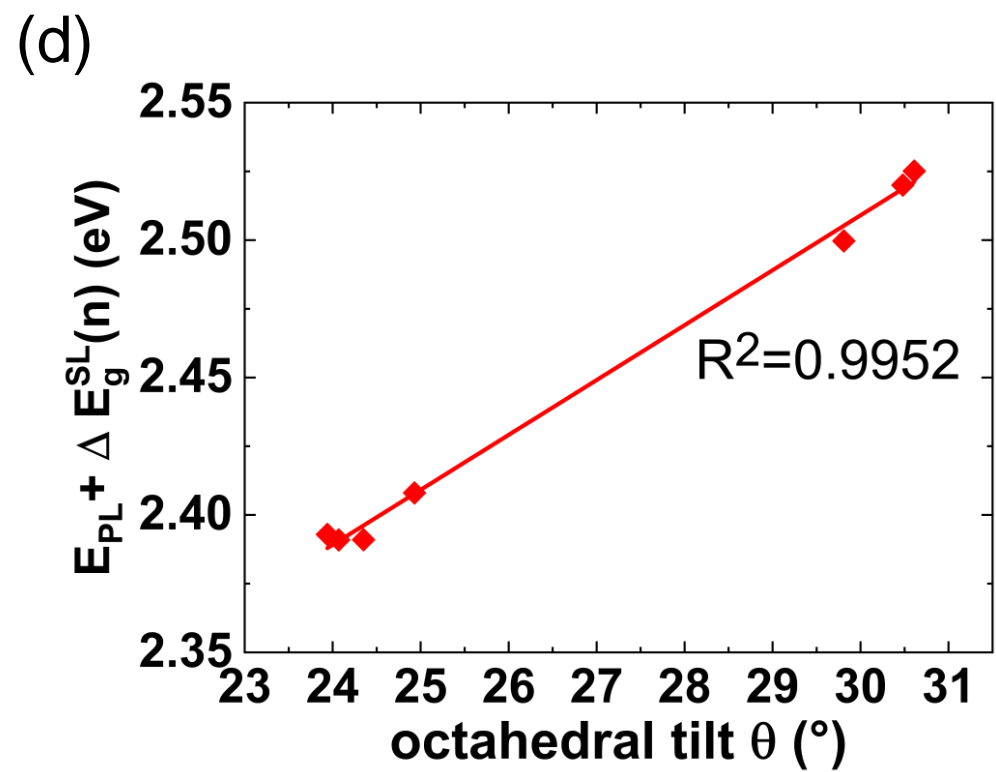